\documentclass[conference]{IEEEtran}
\IEEEoverridecommandlockouts
\usepackage{cite}
\usepackage{amsmath,amssymb,amsfonts}
\usepackage{algorithm}
\usepackage{algpseudocode}
\usepackage{graphicx}
\usepackage{subcaption}
\usepackage{tabularx}
\usepackage{array}
\usepackage{multirow}
\usepackage{makecell}
\usepackage[hidelinks]{hyperref}

\begin{document}

\title{Combinatorial Lock for Trojan-Resilient AI Engines via Hash-Assisted Spectral Filtering}

\author{\IEEEauthorblockN{Soumita Chatterjee, Sudip Ghosh, Tamal Ghosh, Bhargab B. Bhattacharya and Hafizur Rahaman}
\IEEEauthorblockA{Indian Institute of Engineering Science and Technology, Shibpur, India}}

\maketitle

\begin{abstract}
Multi-trigger backdoor attacks, in which heterogeneous trigger mechanisms (e.g., patch, blend, warping, and the latent-domain Spy-Watermark) coexist within a single poisoned model, cannot be addressed by existing shortcut-based defenses that assume a single, static attack mechanism. Current multi-trigger remedies (BU+FT, BU+ANP, and BU+ABL) rely on backdoor unlearning, which incurs high training cost and loses effectiveness at low poisoning ratios. This paper proposes \textbf{Combinatorial Lock}, a training-free, reference-calibrated defense that reduces multi-trigger identification to a combinatorial problem over binary activation signatures. This formulation is the key novelty of the paper: it yields a multi-trigger defense that is training-free, trigger-oblivious, and directly implementable in hardware, since its decision rules can be evaluated with integer and logic operations. The framework combines anomaly-gated spectral scrubbing, which removes the malicious subspace only from inputs identified as anomalous, with a hash-compressed combinatorial signature verified by an ensemble of detection experts, including a compound-attack-aware expert for multi-trigger inputs. A joint threshold--shrinkage search identifies an operating point ($\tau$ at the 97.1st percentile, $\gamma=0.78$) that reduces the attack success rate (ASR) to as low as \textbf{0.1\%} while preserving \textbf{97.8\%} clean accuracy. Evaluation on six models---four CNNs (ResNet18, VGG16, MobileNet, AlexNet) and two vision transformers (ViT-Small, ViT-Base)---shows strong generalization: near-\textbf{100\%} ASR drops to single digits across poisoning rates of \textbf{1--10\%} and \textbf{1--9} simultaneous triggers, with only gradual degradation in clean accuracy. ASIC and FPGA implementations confirm practicality, adding a consistent \textbf{19--20\%} latency overhead across all evaluated architectures and establishing Combinatorial Lock as a hardware-native approach to multi-trigger backdoor mitigation.
\end{abstract}


\section{Introduction}\label{sec:intro}
Advances in artificial intelligence (AI) are driving the deployment of deep neural networks (DNNs) on specialized AI inference hardware. The growing use of AI accelerators in safety-critical applications has widened the attack surface and threatens the robustness of these models \cite{rahaman2024secure,rahaman2024samurai}. Adversaries can exploit attack vectors ranging from data poisoning, adversarial examples \cite{nafi2025dash,rahaman2025runtime}, model extraction, and evasion attacks to hardware-level manipulations that alter model behavior or leak sensitive information \cite{rakin2019bit,rahaman2025runtime}. Among these, backdoor (Trojan) attacks are considered one of the most critical threats: poisoned training samples carry hidden triggers that corrupt inference while the model maintains high accuracy on clean inputs. A further emerging risk is the use of autonomous frameworks for data generation and augmentation, which enable agent-assisted data poisoning during dataset preparation. Current defenses mainly rely on retraining, trigger reconstruction, or poisoned-sample detection; they require significant computational resources and are difficult to integrate into practical hardware systems \cite{rahaman2025secure}.

\begin{figure}[t]
    \centering
    \includegraphics[width=0.95\linewidth]{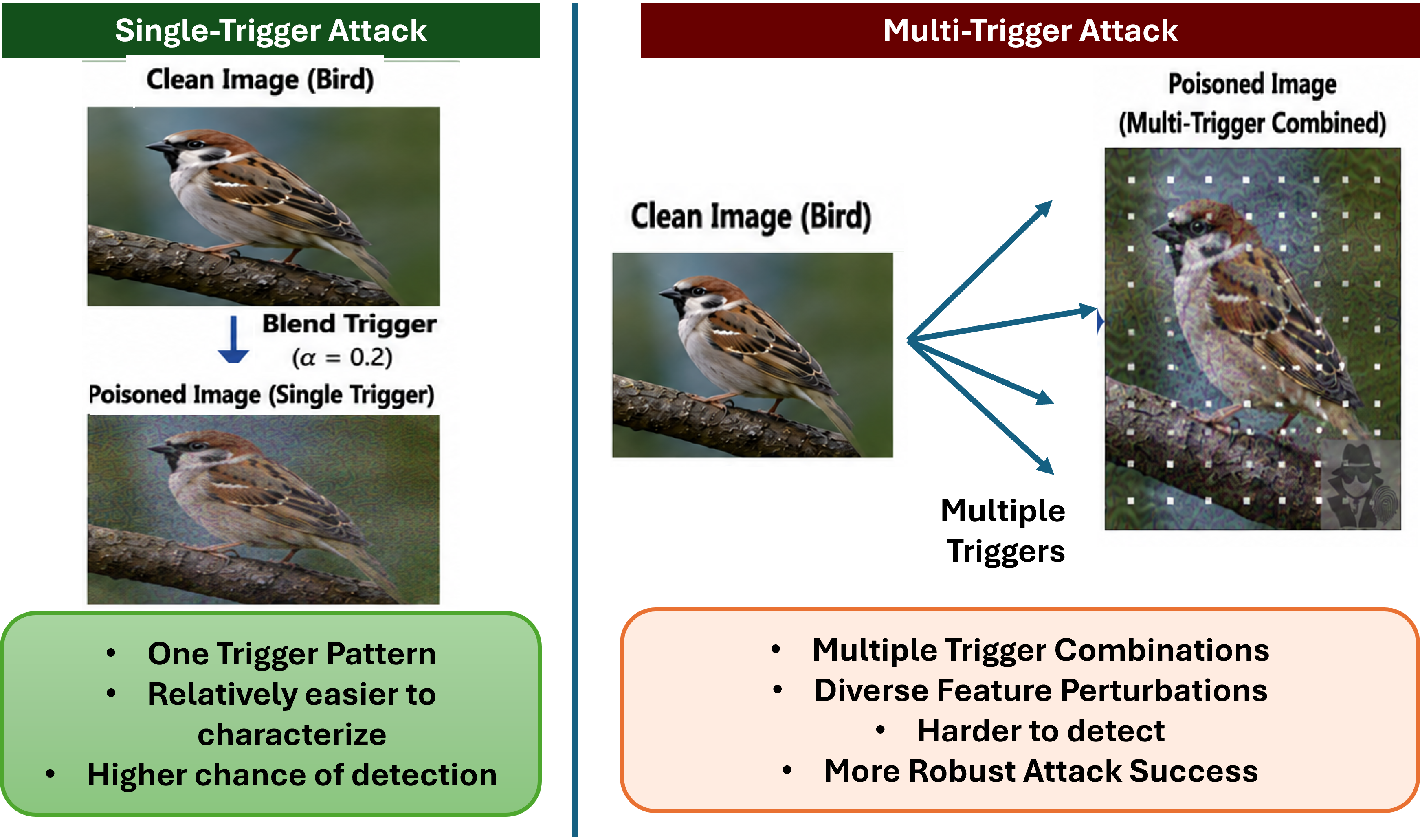}
    \caption{Comparison of single- and multi-trigger backdoors: combining triggers increases attack complexity, diversity, robustness, and resistance to detection.}
    \label{fig:motiv_vlsi}
\end{figure}

A wide variety of triggers has been proposed. BATT~\cite{10096034} uses spatial transformations (translation, rotation) rather than pixel-level perturbations; it activates only under predefined transformations and is effective in both digital and physical settings. The work in~\cite{10134794} studies physical triggers using a facial recognition dataset that combines clean and physically triggered images across multiple trigger patterns. Spy-Watermark~\cite{10448363} embeds a learnable invisible watermark in the latent domain and uses trigger extraction and anti-collapse operations to remain resilient against corruptions and defenses, outperforming ten state-of-the-art attacks on CIFAR-10, GTSRB, and ImageNet. The method in~\cite{10.1109/TDSC.2020.3021407} generates covert, scattered triggers through steganographic embedding and Trojan regularization, evaluated on MNIST, CIFAR-10/100, and GTSRB using ASR and perceptual invisibility (PASS, LPIPS). The work in~\cite{11149824} proposes a watermarking-based invisible backdoor for diffusion-based image editing that preserves normal behavior on clean images while triggering malicious outputs only on watermarked inputs. The work in~\cite{10222807} introduces an any-target backdoor with invisible, input-dependent triggers produced by a steganography model, jointly training the encoder and classifier with pseudo-poisoned samples to strengthen the label--trigger association.

As shown in Fig.~\ref{fig:motiv_vlsi}, multi-trigger backdoor attacks (MTBAs)~\cite{11150583} extend this threat further with sequential, parallel, and hybrid poisoning strategies that let multiple triggers overwrite, coexist with, or cross-activate one another within a model. The work in~\cite{10568200} proposes an M-to-N paradigm that attacks multiple classes simultaneously via invisible, clean-image-based triggers, where each target class can be activated by any of several triggers.

Early defenses centered on trigger reconstruction and anomaly detection. The work in~\cite{8835365} introduced the first trigger reverse-engineering framework, but it requires a clean training set and scales poorly to models with many classes. Subsequent methods~\cite{3454287.3455543,du2019robustanomalydetectionbackdoor} relaxed these requirements through model inversion, conditional generators, and trigger modeling, while~\cite{guo2019taborhighlyaccurateapproach} framed Trojan detection as an optimization problem. These methods, however, remain computationally costly, depend on retraining or clean data, and generalize poorly across diverse triggers. For MTBAs specifically,~\cite{11150583} proposes three remediation strategies---BU+FT, BU+ANP, and BU+ABL---that combine backdoor unlearning with fine-tuning, neuron pruning, and anti-backdoor learning, respectively. All three rely on retraining, which raises computational cost, and they lose effectiveness at low poisoning rates while degrading clean accuracy.

Because existing defenses rely on retraining, trigger reconstruction, or poisoned-sample identification, their practicality for hardware deployment is limited. This work proposes a \textit{Combinatorial Lock} architecture for runtime protection against multi-trigger backdoor attacks. The framework exploits the observation that trigger-induced activations occupy a distinct latent subspace, which enables lightweight hardware verification through spectral filtering and compact binary signatures.

The main contributions are summarized as follows:
\begin{itemize}
\item A training-free, reference-calibrated runtime verification framework that combines anomaly-gated spectral scrubbing with hash-compressed combinatorial signatures, eliminating the need for retraining or a golden model.
\item A two-stage, anomaly-gated defense that filters inputs and scrubs activations to suppress attacks while preserving benign accuracy.
\item A compound-attack-aware ensemble detector evaluated over all $2^8-1=255$ trigger combinations, providing multi-trigger robustness well beyond conventional single-trigger evaluations.
\item A hardware-efficient verification architecture (comparator, XOR, and popcount logic) synthesized on ASIC and FPGA for six DNNs, adding only 19--20\% latency while retaining 97.8\% clean accuracy (CA) at the selected operating point and reducing the attack success rate (ASR) to as low as 0.1\%.
\end{itemize}

The remainder of this paper is organized as follows. Section~\ref{sec:threat} defines the threat model, Section~\ref{sec:method} presents the proposed framework, Section~\ref{sec:results} discusses the experimental results, and Section~\ref{sec:conclusion} concludes the paper with future research directions.

\section{Threat Model}\label{sec:threat}
We consider a frozen, deployed classifier $f$ that has been compromised through training-time data poisoning with multiple backdoor trigger mechanisms. Rather than relying on a single trigger, the attacker employs $K$ heterogeneous trigger patterns~\cite{gu2019badnetsidentifyingvulnerabilitiesmachine, tran2018spectralsignaturesbackdoorattacks, turner2019cleanlabel, chen2017targetedbackdoorattacksdeep, 8802997, khaddaj2023rethinkingbackdoorattacks, liu2018trojaning, nguyen2021wanetimperceptiblewarpingbased, 10448363}, all mapped to the same predefined target label $y_t$. This all-to-one multi-trigger backdoor attack (MTBA) creates hidden shortcuts from every source class to the target class. Among these, Trigger~9---the watermark-based Spy-Watermark~\cite{10448363}, which operates in the latent image domain rather than as a visible pixel-space perturbation---is particularly difficult for prior defenses to mitigate, because its imperceptibility evades detectors calibrated on visible or structurally regular trigger patterns. Our formulation, being trigger-agnostic by construction, remains effective against it.

\begin{figure}[t]
    \centering
    \includegraphics[width=0.95\linewidth]{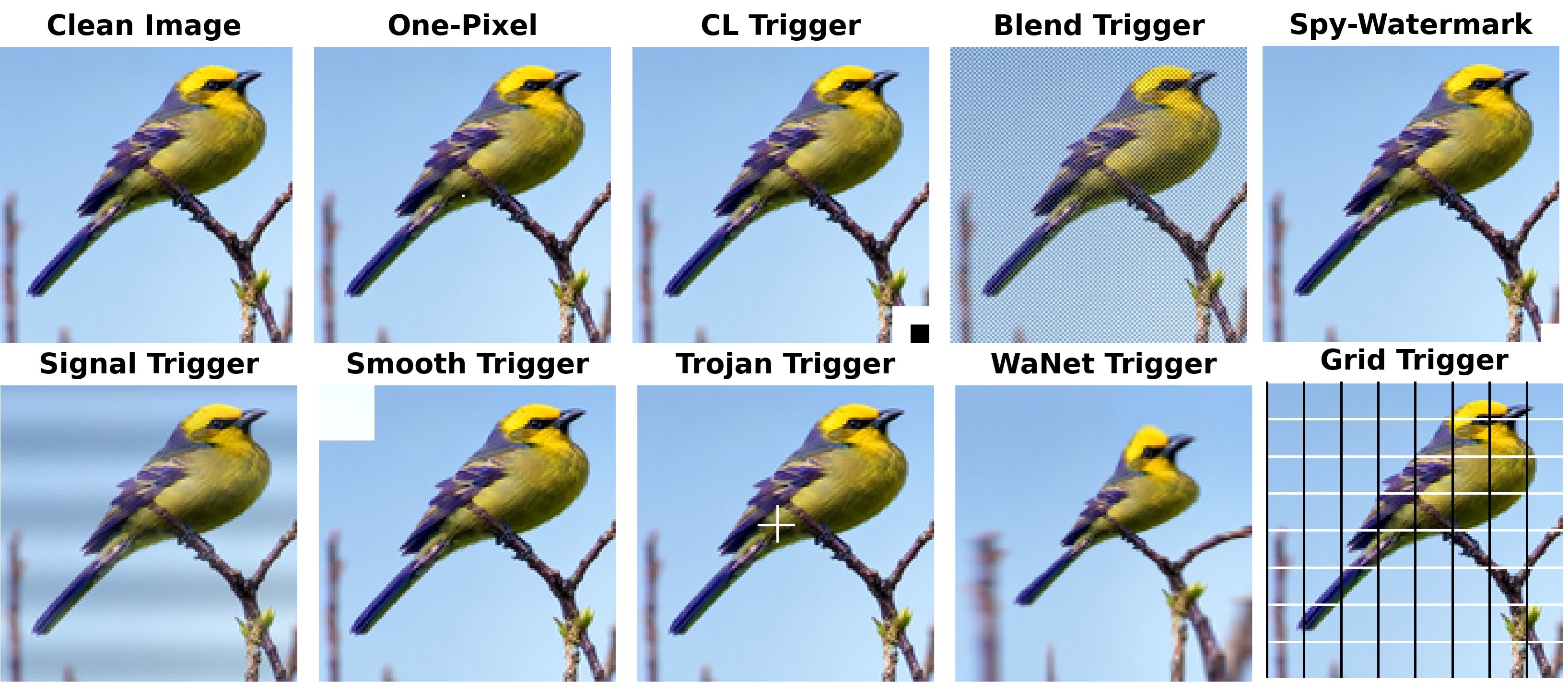}
    \caption{Representative clean, single-trigger, and combined multi-trigger poisoned images used to evaluate the proposed defense.}
    \label{fig:backdoor_trigger}
\end{figure}

Let $\mathcal{D}_c=\{(x_i,y_i)\}_{i=1}^{N_c}$ denote the clean training dataset. The adversary defines $K$ distinct trigger transformations $\{\mathcal{T}_1,\ldots,\mathcal{T}_K\}$, where $\mathcal{T}_k(\cdot)$ embeds the $k$-th trigger into an input image, so that a clean sample $x$ becomes the poisoned sample
\begin{equation}
\tilde{x} = \mathcal{T}_k(x), \qquad k \in \{1,\ldots,K\}.
\end{equation}
For the $k$-th trigger, the adversary selects a subset $\mathcal{P}_k\subset\mathcal{D}_c$, whose size is set by the poisoning rate, and constructs the poisoned subset
\begin{equation}
\mathcal{D}_{b}^{(k)}
=
\left\{
\left(\mathcal{T}_k(x_i),\, y_t\right)
\,\middle|\,
(x_i,y_i) \in \mathcal{P}_k
\right\}.
\end{equation}
The complete poisoned subset is the union of all trigger-specific subsets,
\begin{equation}
\mathcal{D}_b = \bigcup_{k=1}^{K} \mathcal{D}_{b}^{(k)},
\end{equation}
and the training dataset used to train the compromised model is
\begin{equation}
\mathcal{D} = \mathcal{D}_c \cup \mathcal{D}_b.
\end{equation}
After training, the infected classifier behaves normally on clean samples,
\begin{equation}
f(x)=y, \qquad (x,y)\in\mathcal{D}_c,
\end{equation}
while any input that contains one of the embedded triggers is redirected to the target class,
\begin{equation}
f\!\left(\mathcal{T}_k(x)\right) = y_t, \qquad \forall\, k\in\{1,\ldots,K\}.
\end{equation}
The adversary thus establishes multiple independent trigger-to-target mappings within the same model, constituting an all-to-one multi-trigger backdoor attack. Fig.~\ref{fig:backdoor_trigger} shows representative images produced by the triggers used to poison the dataset.

\begin{figure*}[t]
    \centering
    \includegraphics[width=\textwidth]{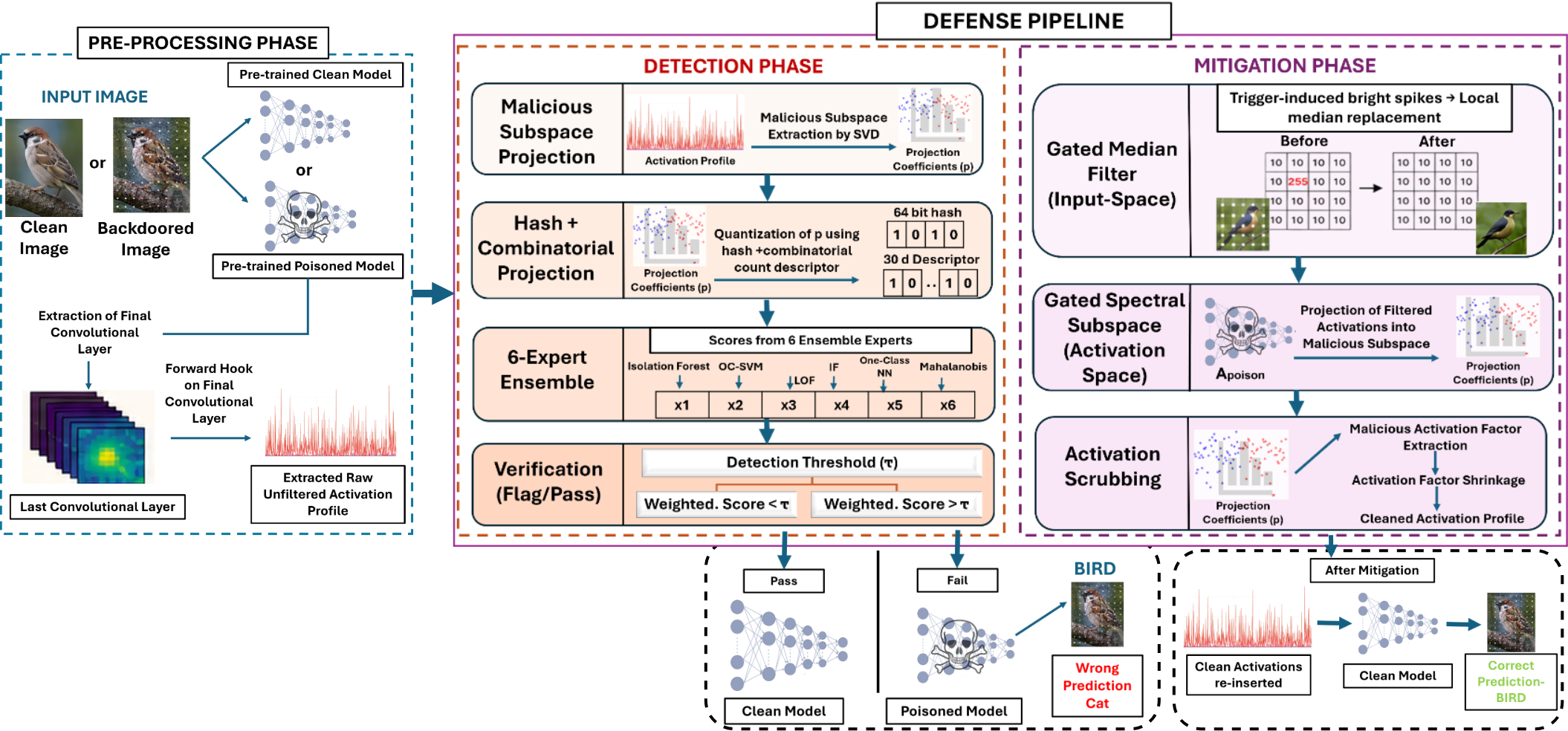}
    \caption{Block diagram of the proposed methodology.}
    \label{block_diagram}
\end{figure*}

\section{Proposed Framework}\label{sec:method}
Deep neural networks learn latent representations that capture the semantic structure of clean inputs. Backdoor poisoning, however, embeds trigger-specific representations that stay dormant on benign samples yet dominate the latent space once a trigger is present. This yields high CA and high ASR at the same time, but it also produces a distinguishable malicious subspace. The proposed framework identifies this subspace, projects inference-time activations onto it, and extracts compact binary descriptors for runtime detection and mitigation.

Fig.~\ref{block_diagram} shows the full pipeline. Final-layer activations feed parallel detection and mitigation branches: detection proceeds through subspace projection, hash descriptors, ensemble analysis, and runtime verification, while mitigation applies gated median filtering and spectral scrubbing to suppress trigger features and preserve benign ones.

\subsection{Feature Extraction and Trigger Validation}\label{sec:features}
In the pre-processing phase, a forward hook is registered on the final feature layer of the model (the last convolutional layer for CNNs) to extract raw, unfiltered activation vectors for every input, whether clean or poisoned. To establish the basis of the defense, this phase includes a strict calibration step in which each selected trigger mechanism is applied to a set of clean calibration images and its raw, undefended ASR is measured.

\subsection{Detection Phase}
\subsubsection{Malicious Subspace Projection}
To separate trigger-induced perturbations from benign activations, singular value decomposition (SVD) is applied to the activation matrix $\mathbf{A} \in \mathbb{R}^{m \times d}$ of the target layer, which holds the activations of $m$ samples over $d$ features:
\begin{equation}
\mathbf{A} = \mathbf{U}\boldsymbol{\Sigma} \mathbf{V}^{\top}.
\end{equation}
Here, $\mathbf{U}$ gives the coordinates of the samples in the latent basis, and triggered samples align more strongly with the leading singular directions than clean samples do. $\boldsymbol{\Sigma}=\mathrm{diag}(\sigma_1,\ldots,\sigma_q)$ contains the singular values, which measure the variance along each singular direction; backdoor triggers depend on a few dominant components. The right singular vectors $\mathbf{v}_i$ satisfy
\begin{equation}
\mathbf{A}^{\top}\mathbf{A}\,\mathbf{v}_i=\lambda_i \mathbf{v}_i,\qquad \lambda_i=\sigma_i^2,
\end{equation}
and form an orthonormal basis of the activation space. Because backdoor-induced activations exhibit large variance along the leading singular directions, the first $r$ right singular vectors, collected in $\mathbf{V}_r\in\mathbb{R}^{d\times r}$, span the malicious subspace. Projecting an activation $\mathbf{z}\in\mathbb{R}^{d}$ onto this subspace yields the projected activation
\begin{equation}
\mathbf{p} = \mathbf{V}_r\mathbf{V}_r^{\top}\mathbf{z} \in \mathbb{R}^{d}.
\end{equation}

\subsubsection{Hash-Guided Feature Projection}
The projected activation $\mathbf{p}$ is transformed into a compact representation for efficient storage and comparison. First, each coefficient is quantized to produce a binary hash,
\begin{equation}
h_i=
\begin{cases}
1, & p_i>\theta_i,\\
0, & \text{otherwise},
\end{cases}
\qquad i=1,\ldots,64,
\end{equation}
where $\theta_i$ denotes the quantization threshold for the $i$-th coefficient. The resulting binary vector $\mathbf{h}\in\{0,1\}^{64}$ is a compact fingerprint of the projected activation.

To preserve higher-order relationships among the hash bits, a complementary combinatorial descriptor is computed as
\begin{equation}
c_j=
\sum_{S\in\mathcal{S}_j}
\mathbf{1}\!\left(\bigwedge_{i\in S} h_i=1\right),
\qquad j=1,\ldots,30,
\end{equation}
where $\mathcal{S}_j$ denotes a predefined collection of bit subsets and $\mathbf{1}(\cdot)$ is the indicator function. The resulting descriptor $\mathbf{c}\in\mathbb{Z}_{\geq 0}^{30}$ captures the co-occurrence statistics of activated bits.

The final feature representation concatenates the binary hash and the combinatorial descriptor,
\begin{equation}
\mathbf{f}=[\mathbf{h};\mathbf{c}] \in \mathbb{Z}_{\geq 0}^{94},
\end{equation}
combining the compactness of binary hashing with the discriminative power of higher-order bit interactions. This hybrid representation is the input to the subsequent multi-expert verification stage.

\begin{figure}[t]
    \centering
    \includegraphics[width=0.95\linewidth]{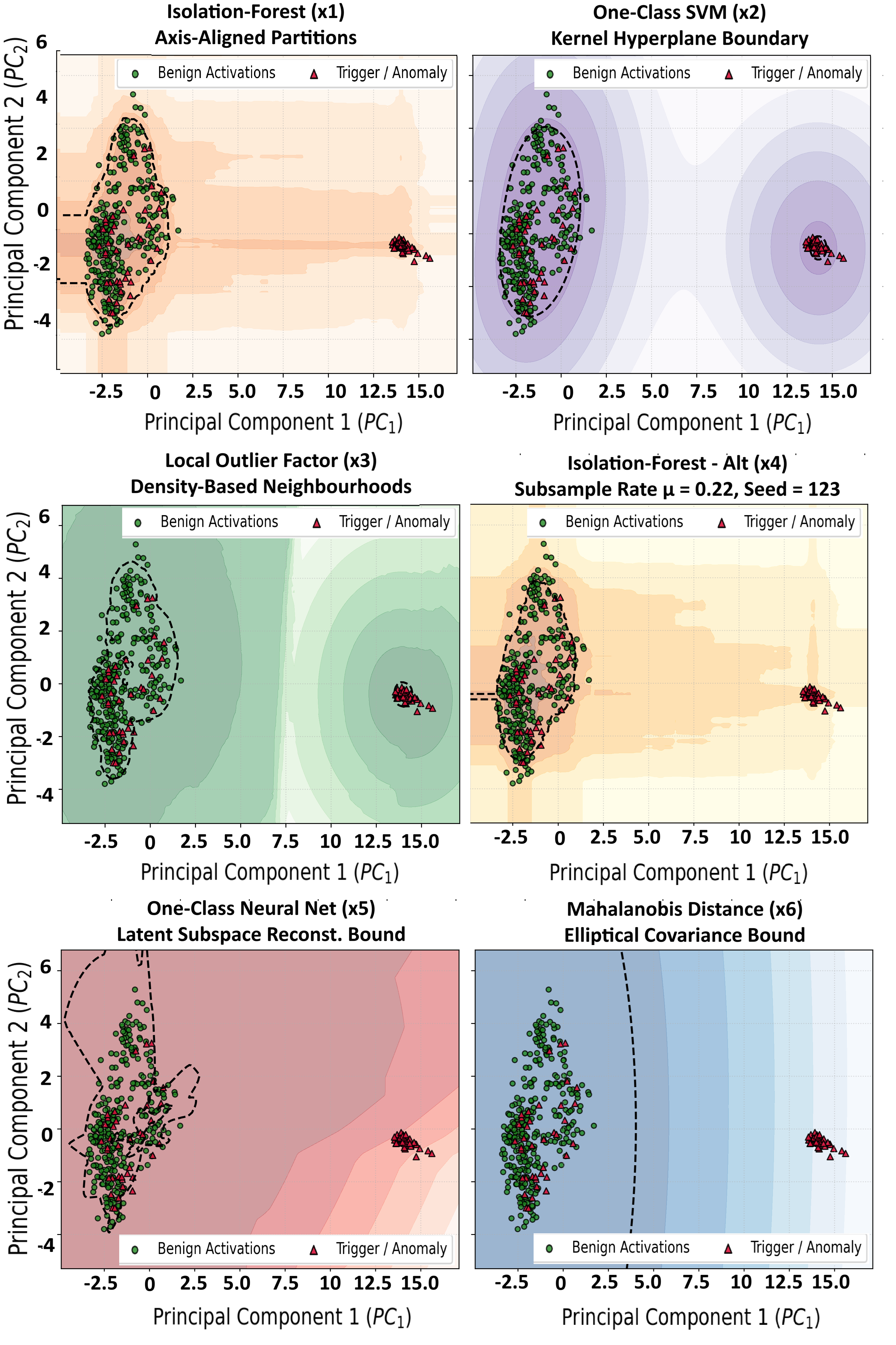}
    \caption{Activation-space decision boundaries of the six base detectors used for anomaly detection.}
    \label{fig:ensemble}
\end{figure}

\subsubsection{Six-Expert Ensemble}
Once an activation has been distilled into its combinatorial hash descriptor, the pipeline must decide whether it reflects benign or malicious behavior. Relying on any single detector for this decision is risky, because each detector embodies its own notion of ``normal.'' Fig.~\ref{fig:ensemble} overlays the decision boundaries of the ensemble's six base detectors on a 2D PCA projection (benign samples in green, anomalies in red), revealing both a well-separated anomaly cluster and subtler anomalies intermixed with benign points. The detectors---including Isolation Forest, One-Class SVM, Local Outlier Factor (LOF), and Mahalanobis distance, which span axis-aligned, kernel, density, and covariance-based geometries---all capture the separated cluster, but only some of them catch the intermixed cases, which explains why no single geometry suffices. Combining all six into one weighted ensemble score closes these blind spots and makes simultaneous evasion of all detectors far harder.

\subsubsection{Verification Phase}
The six anomaly scores produced by the expert ensemble are aggregated into a single verification score by a weighted linear combination,
\begin{equation}
S=\sum_{i=1}^{6}w_i s_i,
\qquad
\sum_{i=1}^{6}w_i=1,
\end{equation}
where $s_i$ denotes the anomaly score of the $i$-th expert and $w_i$ its reliability weight. The weights are estimated on a validation set so that more discriminative experts contribute proportionally more to the final decision. The aggregated score is compared with a detection threshold $\tau$ to accept or reject the input:
\begin{equation}
\mathrm{Decision}=
\begin{cases}
\mathrm{Pass}, & S < \tau,\\
\mathrm{Flag}, & S \geq \tau.
\end{cases}
\end{equation}
Unlike simple majority voting, weighted fusion integrates the complementary evidence of the experts into a single confidence value, improving robustness against detectors that fail on specific triggers. The value of $\tau$ is selected on the validation set to obtain the required balance between detection accuracy and false positives.

\begin{algorithm}[t]
\caption{Three-Stage Mitigation Framework}
\label{alg:mitigation}
\begin{algorithmic}[1]
\Require Input image $x$; deployed classifier $f=g\circ\phi$ with feature extractor $\phi$ and classification head $g$; malicious subspace basis $\mathbf{V}_r\in\mathbb{R}^{d\times r}$; median threshold $\tau_m$; shrinkage factor $\gamma$
\Ensure Predicted label $\hat{y}$
\Statex \textbf{Stage 1: Gated Median Filter (Input Space)}
\State Detect trigger-induced bright spikes in $x$
\If{any pixel intensity $>\tau_m$}
    \State Replace each outlier pixel with its local neighborhood median
\EndIf
\State Obtain the filtered image $x_f$
\Statex \textbf{Stage 2: Gated Spectral Subspace Projection}
\State Extract deep feature activations: $\mathbf{z} \gets \phi(x_f)$
\State Project onto the malicious subspace: $\mathbf{p} \gets \mathbf{V}_r\mathbf{V}_r^{\top}\mathbf{z}$
\Statex \textbf{Stage 3: Activation Scrubbing}
\State Suppress the malicious component by spectral shrinkage: $\mathbf{z}_{\mathrm{clean}} \gets \mathbf{z}-\gamma\mathbf{p}$
\State Predict the output label: $\hat{y} \gets g(\mathbf{z}_{\mathrm{clean}})$
\State \Return $\hat{y}$
\end{algorithmic}
\end{algorithm}

\subsection{Mitigation Phase}\label{sec:mitigation}
Algorithm~\ref{alg:mitigation} summarizes the three-stage mitigation pipeline, which sanitizes both the input space and the internal representations before classification. First, the input $x$ passes through a gated median filter: pixels exceeding the median threshold $\tau_m$, which indicates a trigger-induced bright spike, are replaced by their neighborhood median, yielding the sanitized image $x_f$. Second, the deep feature activations $\mathbf{z} = \phi(x_f)$ are extracted and projected onto the precomputed malicious subspace basis $\mathbf{V}_r$ to obtain $\mathbf{p} = \mathbf{V}_r\mathbf{V}_r^{\top}\mathbf{z}$, the component of $\mathbf{z}$ that lies in the malicious subspace. Third, this component is removed by spectral shrinkage with factor $\gamma$, $\mathbf{z}_{\mathrm{clean}} = \mathbf{z} - \gamma \mathbf{p}$. Finally, the sanitized representation is passed to the classification head, $\hat{y}=g(\mathbf{z}_{\mathrm{clean}})$.

\section{Results and Discussion}\label{sec:results}
\subsection{Combinatorial Lock and Mitigation Parameter Analysis}

\begin{figure*}[t]
    \centering
    \begin{subfigure}[b]{0.30\textwidth}
        \centering
        \includegraphics[width=\textwidth]{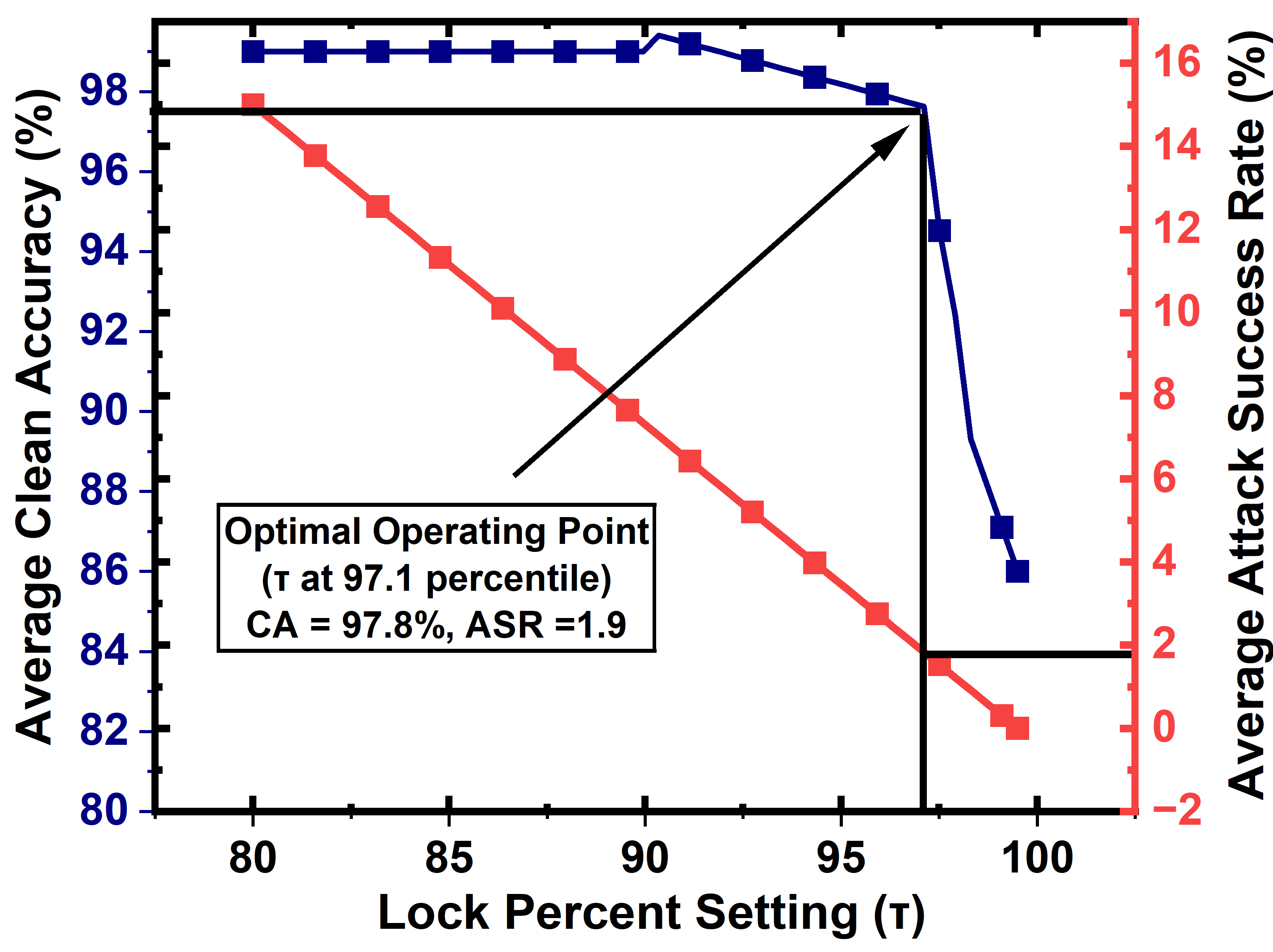}
        \caption{Validation of $\tau$}
        \label{fig:sense_1}
    \end{subfigure}
    \hfill
    \begin{subfigure}[b]{0.34\textwidth}
        \centering
        \includegraphics[width=\textwidth]{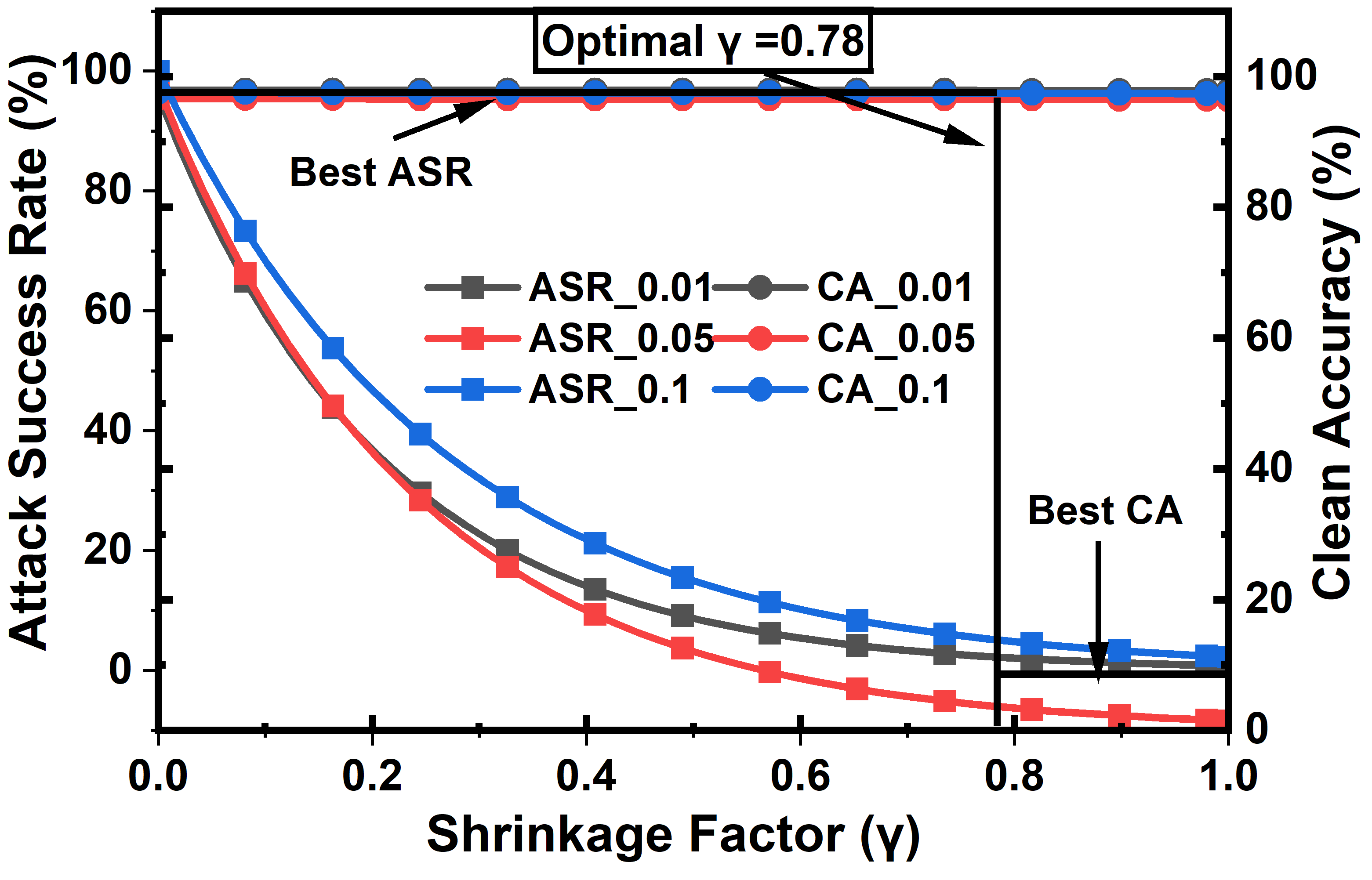}
        \caption{Validation of $\gamma$}
        \label{fig:sense_2}
    \end{subfigure}
    \hfill
    \begin{subfigure}[b]{0.30\textwidth}
        \centering
        \includegraphics[width=\textwidth]{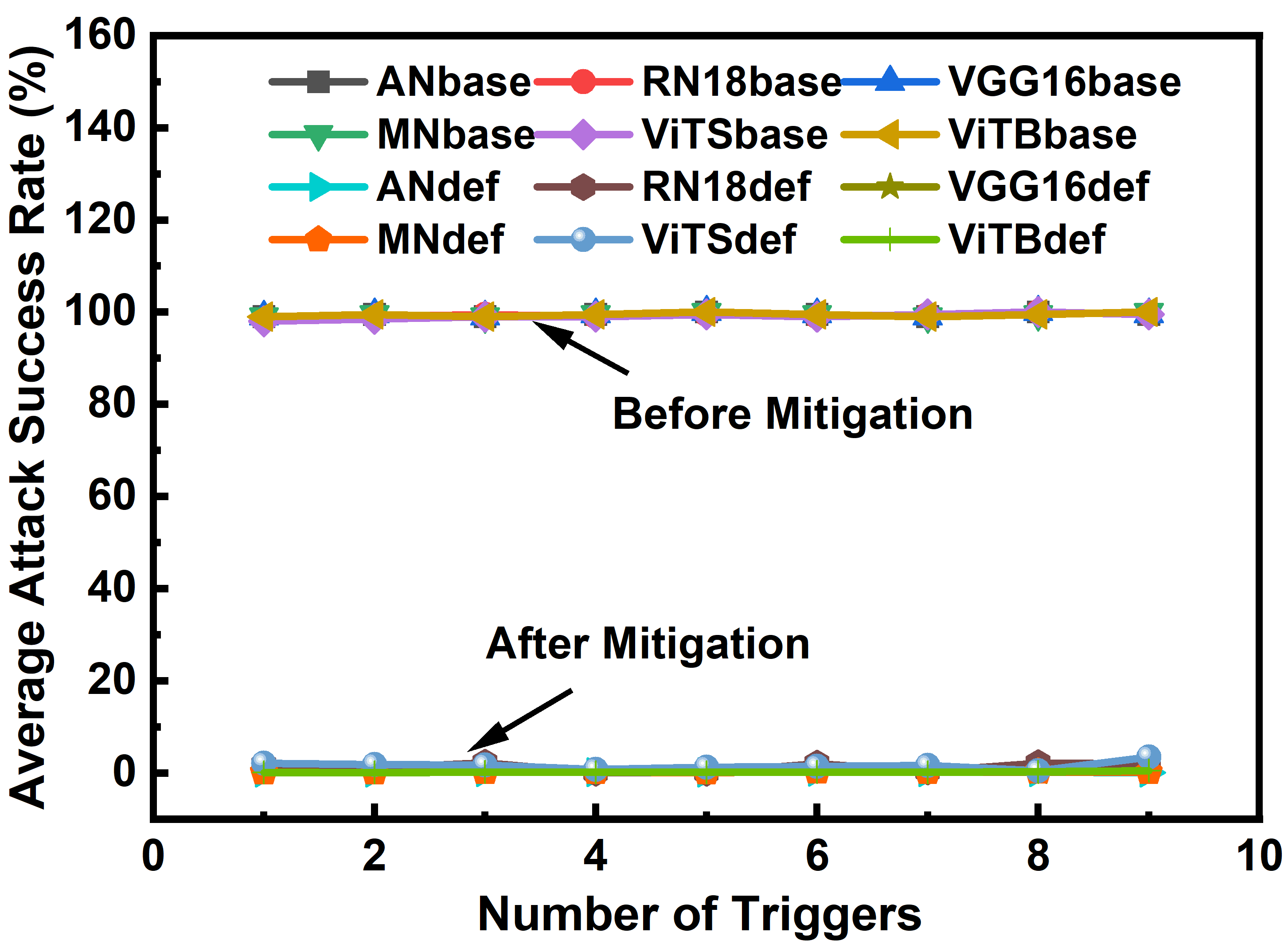}
        \caption{Average ASR across poisoning rates}
        \label{fig:asr}
    \end{subfigure}
    \caption{Parameter sensitivity and ASR robustness of Combinatorial Lock.}
    \label{fig:metric_comparison}
\end{figure*}

\begin{table*}[t]
\centering
\caption{Clean accuracy (CA, \%) across architectures, poisoning rates, and trigger counts}
\label{tab:defense_comparison}
\renewcommand{\arraystretch}{1.15}
\setlength{\tabcolsep}{4pt}
\resizebox{\textwidth}{!}{%
\begin{tabular}{|c|c|c|c|c|c|c|c|c|c|c|c|c|c|c|c|c|c|c|}
\hline
\multirow{2}{*}{\makecell{\textbf{Trigger}\\\textbf{Count}}}
& \multicolumn{3}{c|}{\textbf{ResNet18}}
& \multicolumn{3}{c|}{\textbf{VGG16}}
& \multicolumn{3}{c|}{\textbf{MobileNet}}
& \multicolumn{3}{c|}{\textbf{AlexNet}}
& \multicolumn{3}{c|}{\textbf{ViT-Small}}
& \multicolumn{3}{c|}{\textbf{ViT-Base}} \\
\cline{2-19}
& \textbf{1\%} & \textbf{5\%} & \textbf{10\%}
& \textbf{1\%} & \textbf{5\%} & \textbf{10\%}
& \textbf{1\%} & \textbf{5\%} & \textbf{10\%}
& \textbf{1\%} & \textbf{5\%} & \textbf{10\%}
& \textbf{1\%} & \textbf{5\%} & \textbf{10\%}
& \textbf{1\%} & \textbf{5\%} & \textbf{10\%} \\
\hline
1 & 98.60 & 98.58 & 98.56 & 94.36 & 94.28 & 94.22 & 92.48 & 92.39 & 92.42 & 91.58 & 90.35 & 89.95 & 97.82 & 95.02 & 89.20 & 93.82 & 92.71 & 90.05 \\
\hline
2 & 98.52 & 98.46 & 98.50 & 94.23 & 94.08 & 94.09 & 92.35 & 92.12 & 92.31 & 91.40 & 90.13 & 89.68 & 97.66 & 94.53 & 88.98 & 93.65 & 92.54 & 89.86 \\
\hline
3 & 98.30 & 98.54 & 98.43 & 94.10 & 93.92 & 94.09 & 92.01 & 92.24 & 92.27 & 91.08 & 89.97 & 89.64 & 97.42 & 93.75 & 88.85 & 93.45 & 92.30 & 89.25 \\
\hline
4 & 98.42 & 98.37 & 98.41 & 93.70 & 93.94 & 93.92 & 91.84 & 91.78 & 91.90 & 90.81 & 89.41 & 89.42 & 97.68 & 92.97 & 88.70 & 92.97 & 92.28 & 89.06 \\
\hline
5 & 98.17 & 98.27 & 98.34 & 93.76 & 93.70 & 93.68 & 91.78 & 91.79 & 91.82 & 90.69 & 89.30 & 89.50 & 96.88 & 95.31 & 88.50 & 92.50 & 91.89 & 89.00 \\
\hline
6 & 98.19 & 98.24 & 98.09 & 93.29 & 93.43 & 93.64 & 91.76 & 91.59 & 91.65 & 90.63 & 89.12 & 88.83 & 96.80 & 94.89 & 88.32 & 92.48 & 91.45 & 88.89 \\
\hline
7 & 97.97 & 98.05 & 98.14 & 93.23 & 93.14 & 93.35 & 91.30 & 91.45 & 91.26 & 90.47 & 88.60 & 89.09 & 97.60 & 94.66 & 88.12 & 92.10 & 91.23 & 88.20 \\
\hline
8 & 97.98 & 97.96 & 98.00 & 93.11 & 93.16 & 93.16 & 91.29 & 91.10 & 91.31 & 90.23 & 88.47 & 88.44 & 96.88 & 95.20 & 87.50 & 91.96 & 90.89 & 87.50 \\
\hline
9 & 97.84 & 97.80 & 97.87 & 92.98 & 93.01 & 93.04 & 91.14 & 90.98 & 91.12 & 90.08 & 88.31 & 88.26 & 96.62 & 94.92 & 87.32 & 91.74 & 90.63 & 87.31 \\
\hline
\end{tabular}%
}
\end{table*}

\begin{table*}[t]
\centering
\caption{ASIC and FPGA implementation results of the proposed defense across DNN architectures}
\label{tab:hardware_results}
\renewcommand{\arraystretch}{1.15}
\resizebox{\textwidth}{!}{%
\begin{tabular}{|l|c|c|c|c|c|c|c|c|c|c|c|c|}
\hline
\textbf{Model} & \textbf{Cells} & \textbf{Area} & \textbf{Power} & \textbf{ASIC} & \textbf{FFs} & \textbf{LUTs} & \textbf{FPGA} & \textbf{FP32} & \textbf{INT8} & \textbf{Base} & \textbf{Defended} & \textbf{Overhead} \\
& \textbf{(ASIC)} & \textbf{($\mu$m$^2$)} & \textbf{(mW)} & \textbf{Fmax (MHz)} & & & \textbf{Fmax (MHz)} & \textbf{Mem. (KB)} & \textbf{Mem. (KB)} & \textbf{Latency (ms)} & \textbf{Latency (ms)} & \textbf{(\%)} \\
\hline
AlexNet & 3248 & 19241.57 & 9.834 & 218 & 136 & 954 & 185 & 14.82 & 3.71 & 4.3812 & 5.2247 & 19.25 \\
\hline
MobileNet & 3706 & 21918.34 & 10.562 & 210 & 152 & 1068 & 178 & 17.45 & 4.36 & 5.2186 & 6.2188 & 19.17 \\
\hline
ResNet18 & 4161 & 24606.82 & 11.907 & 202 & 168 & 1181 & 171 & 25.23 & 6.31 & 6.5413 & 7.8345 & 19.77 \\
\hline
VGG16 & 6512 & 38964.75 & 18.483 & 186 & 246 & 1734 & 158 & 58.34 & 14.59 & 11.4821 & 13.7645 & 19.88 \\
\hline
ViT-Small & 5374 & 32167.84 & 15.236 & 193 & 212 & 1498 & 164 & 42.71 & 10.68 & 9.2437 & 11.0516 & 19.56 \\
\hline
ViT-Base & 7486 & 44783.56 & 21.347 & 178 & 294 & 2078 & 151 & 86.42 & 21.61 & 15.3268 & 18.3972 & 20.03 \\
\hline
\end{tabular}%
}
\end{table*}

The choice of the lock threshold $\tau$ governs the balance between backdoor suppression and clean-model accuracy. Rather than fixing $\tau$ heuristically, the proposed approach sweeps $\tau$ across its percentile range while measuring CA and ASR, as shown in Fig.~\ref{fig:sense_1}. As $\tau$ increases from the 80th to the 97th percentile, CA decreases only slightly (from 99\% to 97.5\%), whereas ASR drops sharply (from 15\% to 2\%). Beyond the 97.1st percentile, filtering becomes too aggressive for benign activations, causing a considerable drop in CA without any further improvement in ASR. The optimal threshold is therefore the 97.1st percentile, at which CA and ASR are 97.8\% and 1.9\%, respectively.

The use of soft thresholding is motivated by Chebyshev's inequality, which, following the outlier-based rationale of spectral signatures~\cite{tran2018spectralsignaturesbackdoorattacks}, provides a distribution-free bound on large deviations of a latent feature $H$ with mean $\mu$ and standard deviation $\sigma$:
\begin{equation}
P\left(|H-\mu|\geq \kappa\sigma\right)\leq \frac{1}{\kappa^2}.
\end{equation}
Backdoor-induced activations appear as high-variance outliers (typically beyond $3\sigma$) relative to the benign feature distribution, whereas, for $\kappa=3$, at most $1/9$ ($\approx 11\%$) of benign activations can deviate this far, regardless of their distribution. This motivates applying soft thresholding to the dominant $r=18$ singular components rather than removing them entirely. As illustrated in Fig.~\ref{fig:sense_2}, $\gamma=0.78$ provides the best compromise between attack suppression and clean accuracy: smaller values leave residual trigger features, and larger values unnecessarily degrade benign performance.

\subsection{Detector Quality}
To select the number of expert subspaces $k$, we vary $k$ from 1 to 12 and compare the AUROC of the ensemble with two non-ensemble baselines, rule-based magnitude thresholding (AUROC = 0.9376) and joint Mahalanobis distance (AUROC = 0.9666); the results are listed in Table~\ref{tab:ensemble_comparison}. AUROC follows a clearly unimodal trend. It rises from 0.8957 at $k=1$, where a single coarse expert underfits the activation subspace, peaks at $k=5$ and $k=6$ (0.9800 and 0.9854), and then declines as larger values of $k$ split the signal into noise-dominated parts, falling below both baselines for all reported $k\geq 9$. The ensemble outperforms the stronger baseline only for $4\leq k\leq 7$, with the largest margin at $k=6$, whereas poorly sized ensembles ($k\leq 2$ or $k\geq 9$) underperform both baselines.

\begin{table}[t]
\centering
\caption{Comparison of the proposed expert ensemble with baselines}
\label{tab:ensemble_comparison}
\renewcommand{\arraystretch}{1.15}
\small
\begin{tabular}{|l|c|c|}
\hline
\textbf{Method} & \textbf{AUROC} & \makecell{\textbf{$\Delta$ vs.\ Joint}\\\textbf{Mahalanobis (\%)}} \\
\hline
\multicolumn{3}{|c|}{\textbf{Non-Ensemble Baselines}} \\
\hline
Rule-Based Scoring & 0.9376 & $-3.00$ \\
\hline
Joint Mahalanobis & 0.9666 & -- \\
\hline
\multicolumn{3}{|c|}{\textbf{Expert Ensemble ($k$-Sweep)}} \\
\hline
$k=1$ & 0.8957 & $-7.33$ \\
\hline
$k=2$ & 0.9184 & $-4.99$ \\
\hline
$k=4$ & 0.9715 & $+0.51$ \\
\hline
$k=5$ & \textbf{0.9800} & $\mathbf{+1.39}$ \\
\hline
$k=6$ & \textbf{0.9854} & $\mathbf{+1.94}$ \\
\hline
$k=7$ & 0.9711 & $+0.47$ \\
\hline
$k=9$ & 0.9204 & $-4.78$ \\
\hline
$k=10$ & 0.8872 & $-8.21$ \\
\hline
$k=12$ & 0.8403 & $-13.07$ \\
\hline
\end{tabular}
\end{table}

\subsection{Cross-Platform Hardware Evaluation}
The Combinatorial Lock detection and mitigation datapath was synthesized for ASIC with Synopsys Design Compiler in a 90~nm technology and implemented separately on a Xilinx PYNQ UltraScale+ FPGA with Xilinx Vivado, for each of the six model backbones; the results are given in Table~\ref{tab:hardware_results}. Cell count, area, and power grow with the backbone, from 3,248 cells, 19,241.57~$\mu$m$^2$, and 9.834~mW for AlexNet to 7,486 cells, 44,783.56~$\mu$m$^2$, and 21.347~mW for ViT-Base, while the same defense logic is instantiated for every backbone without architecture-specific redesign. INT8 quantization reduces the defense's memory footprint by a consistent factor of $4\times$ relative to FP32 for all models (e.g., from 14.82~KB to 3.71~KB for AlexNet). The latency overhead remains nearly constant, between 19.17\% and 20.03\%, across all six backbones. This near-constant relative cost is what makes the result significant for AI accelerators: because the cost of the defense tracks the cost of the base model rather than imposing a disproportionate additional burden, the protected inference path can be realized directly on ASIC or FPGA inference hardware without a separate security co-processor.

\subsection{Comparative Performance Analysis}
Experiments were run on an NVIDIA RTX A2000 GPU using PyTorch, NumPy, and SciPy. To assess generalization and stability, the defense was evaluated on CIFAR-10 across six architectures---ResNet18, VGG16, MobileNet, AlexNet, ViT-Small, and ViT-Base---under poisoning rates of 1\%, 5\%, and 10\% and trigger counts from 1 to 9.

As Table~\ref{tab:defense_comparison} shows, CA declines only marginally as the poisoning rate and the trigger count increase; for ResNet18, it decreases from 98.60\% (1 trigger, 1\% poisoning) to 97.87\% (9 triggers, 10\% poisoning). ResNet18 achieves the highest CA in every configuration, while the transformer-based models are the most sensitive to the poisoning rate; for example, ViT-Small drops from 97.82\% at 1\% poisoning to 89.20\% at 10\% poisoning with a single trigger.

ASR is measured for all six models at each poisoning rate and averaged over poisoning rates in Fig.~\ref{fig:asr}. Before the defense, ASR remains at 98--100\% for all six models regardless of trigger count, showing how reliably the backdoors succeed even as attack complexity grows. After the defense, ASR falls to roughly 0--2\% across all models and trigger counts, with only a marginal increase for ResNet18 (RN18-def) as the trigger count approaches 9.

Table~\ref{tab:state_comparison} compares the proposed method with BU+FT, BU+ANP, and BU+ABL~\cite{11150583} on ResNet18, VGG16, MobileNet, and AlexNet, with each value averaged over the 1\%, 5\%, and 10\% poisoning rates. The proposed method achieves the highest CA on ResNet18 (98.25\% vs.\ 98.09--98.11\%) and AlexNet (89.81\% vs.\ 88.88--88.91\%), exceeds BU+FT and BU+ABL on VGG16 (93.65\% vs.\ 93.48--93.49\%), and stays within 0.10 percentage points of the best baseline on MobileNet (91.76\% vs.\ 91.86\%). At the same time, it reduces ASR to 0.15--1.27\%, compared with 1.90--4.35\% for the baselines. Whereas the baselines trade some accuracy for security, or vice versa, the proposed approach substantially improves security with little or no cost in accuracy.

\begin{table}[t]
\centering
\caption{Average CA and ASR (\%) across architectures}
\label{tab:state_comparison}
\footnotesize
\setlength{\tabcolsep}{2.5pt}
\begin{tabularx}{\columnwidth}{|l|*{8}{>{\centering\arraybackslash}X|}}
\hline
\multirow{2.5}{*}{\textbf{Method}} & \multicolumn{2}{c|}{\textbf{ResNet18}} & \multicolumn{2}{c|}{\textbf{VGG16}} & \multicolumn{2}{c|}{\textbf{MobileNet}} & \multicolumn{2}{c|}{\textbf{AlexNet}} \\ \cline{2-9}
 & \textbf{CA} & \textbf{ASR} & \textbf{CA} & \textbf{ASR} & \textbf{CA} & \textbf{ASR} & \textbf{CA} & \textbf{ASR} \\ \hline
BU+FT~\cite{11150583}  & 98.11 & 3.20 & 93.49 & 3.75 & 91.86 & 3.85 & 88.88 & 4.35 \\ \hline
BU+ANP~\cite{11150583} & 98.09 & 1.90 & 98.54 & 1.95 & 91.45 & 2.25 & 88.89 & 3.01 \\ \hline
BU+ABL~\cite{11150583} & 98.10 & 2.25 & 93.48 & 2.87 & 91.60 & 2.00 & 88.91 & 2.55 \\ \hline
\textbf{Proposed} & \textbf{98.25} & \textbf{0.15} & \textbf{93.65} & \textbf{0.17} & \textbf{91.76} & \textbf{0.19} & \textbf{89.81} & \textbf{1.27} \\ \hline
\end{tabularx}
\end{table}

\section{Conclusion}\label{sec:conclusion}
This paper presented \textbf{Combinatorial Lock}, a training-free defense against multi-trigger backdoor attacks that distills each activation profile into a lightweight binary runtime signature and reduces detection to purely \emph{combinatorial} operations---Hamming distances, popcounts, and XOR comparisons---using only integer and logic arithmetic. Unlike prior work, this makes the defense training-free, architecture-agnostic, and hardware-native: BU+FT, BU+ANP, and BU+ABL require costly retraining and degrade at low poisoning rates, while spectral and statistical detectors were never designed for silicon realization. Jointly optimizing the lock threshold and the spectral shrinkage factor yields an effective operating point ($\tau$ at the 97.1st percentile, $\gamma=0.78$) that reduces the average ASR to \textbf{1.9\%}, with a lowest ASR of 0.1\%, while preserving \textbf{97.8\%} clean accuracy. Across 1--9 triggers and six architectures (ResNet18, VGG16, MobileNet, AlexNet, ViT-Small, and ViT-Base), near-\textbf{100\%} ASR collapses to single digits. Notably, even the latent-domain Spy-Watermark trigger---imperceptible and structurally dissimilar to conventional pixel-space patterns, and therefore difficult for prior detectors calibrated on visible perturbations---is reliably detected under the combinatorial signature formulation. ASIC and FPGA implementations confirm practicality, adding only \textbf{19--20\%} latency overhead with modest area, power, and memory costs (\textbf{4--22~KB} of INT8 storage), which establishes Combinatorial Lock as a retraining-free, architecture-agnostic, and natively hardware-realizable multi-trigger backdoor defense.

Future work will investigate adaptive threshold selection, robustness against unseen and adaptive triggers, extension to multimodal and foundation models, and hardware--software co-design to further reduce overhead while preserving security guarantees.

\bibliographystyle{IEEEtran}
\bibliography{Arxiv_ref}

\end{document}